\documentclass[aps,prl,twocolumn,showpacs,superscriptaddress,groupedaddress]{revtex4}  
\usepackage{graphicx}  
\usepackage{dcolumn}   
\usepackage{bm}        
\usepackage{amssymb}   

\begin{document}



\title{Vesicle adhesion reveals novel universal relationships for \\ biophysical characterization }







\author{Ehsan Irajizad, Ashutosh Agrawal$^*$\\
{\em  Department of Mechanical Engineering, University of Houston, Houston, TX, USA}\\
$^*$Corresponding author, e-mail: ashutosh@uh.edu\\
}
\date{\today}

\begin{abstract}
Adhesion plays an integral role in diverse biological functions ranging from cellular transport to tissue development. Estimation of adhesion strength, therefore, becomes important to gain biophysical insight into these phenomena. In this Letter, we use curvature elasticity to present non-intuitive, yet remarkably simple, universal relationships that capture vesicle-substrate interactions. Our study reveals that the inverse of the height, exponential of the contact area, and the force required to detach the vesicle from the substrate vary linearly with the square root of the adhesion energy. These relationships not only provide efficient strategies to tease out adhesion energy of biological molecules but can also be used to characterize the physical properties of elastic biomimetic nanoparticles. We validate the modeling predictions with experimental data from two previous studies. 

\end{abstract}

\maketitle

\begin{figure}[t]
\begin{center} 
\includegraphics[height =40mm, keepaspectratio]{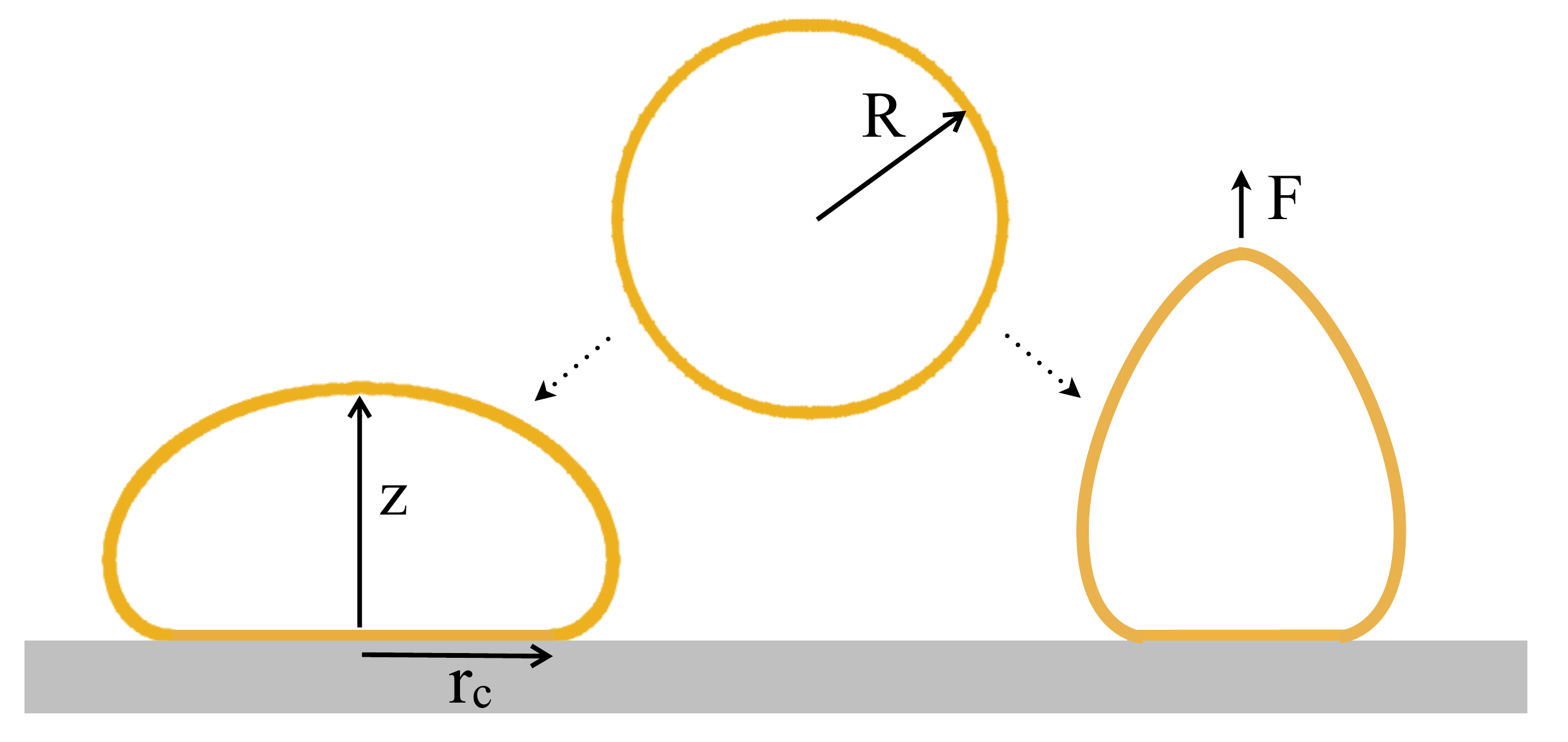} \hspace{0cm} \label{a}
\caption{Vesicle-substrate interactions  yield universal relationships for biomechanical characterization. We model the equilibriated shape of the adhered vesicle on a planar substrate and  compute the vesicle height, contact area, membrane tension and the critical vertical force required to detach the vesicle from the substrate for different adhesion energies.} 
\label{Synergy1}
\vspace{-1 cm}
\end{center}
\end{figure}

Adhesion between biological molecules plays a critical role in cell transport, cell migration, signal transduction, and tissue development\cite{alberts2008molecular,goldstein1979coated,gumbiner1996cell,aplin1998signal,huttenlocher1995adhesion,rutishauser1988cell}. A wide range of proteins that mediate cargo-cell and cell-cell interactions are dedicated for this purpose \cite{alberts2008molecular,edelman1983cell, albelda1990integrins,juliano2002signal}. Adhesion between neuronal cells in the central nervous system has also been implicated in establishing synaptic plasticity and memory\cite{benson2000making}. While the functional relevance of such adhesive interactions is well recognized, characterizing the strength of the adhesion molecules is still a challenging task. A lack of quantification of the strength of these interactions is a deterrent to biophysical investigations of the related cellular processes. In addition, recent advances in material science have led to the design of sophisticated biomimetic and shape-switching nanoparticles for drug delivery and medical diagnostics \cite{Samir2008,Samir2015,samir2009Red,samir2010shape,Mitragotri2011Nature, Bao2015ACSNANO,Zhang2015Nature,Ferrari2008Biomaterial}. As studies have unequivocally highlighted the role of adhesive and elastic properties of nanoparticles in regulating their blood circulation, targeting and cellular uptake, there is a pressing need to characterize their physical properties for achieving optimal design and functionalization \cite{Samir2008,Ferrari2008Biomaterial,Samir2015,Mitragotri2011Nature,Zhang2010PRL,Bao2015ACSNANO}. As will be argued later, Young's modulus measured from existing biophysical techniques is geometry-dependent and is therefore, not the most relevant measure of material property for soft shell-like nanoparticles. In this Letter, we use nonlinear computational modeling to present universal relationships that allow estimation of adhesion energy from the adhesion of vesicles on rigid planar substrates. Remarkably, these same relationships can be exploited to characterize the surface and elastic properties of synthetic nanoparticles. 



Vesicle adhesion has been a subject of active research for almost three and a half decades  \cite{Evans1980,Seifert1990,Lipowsky1991,Seifert1991,Albersdorfer1997,Guttenberg2001,Capovilla2002,Lai2002,Cuvelier2004,Gruhn2007,Lorz2007,Deserno2007,Sengupta2010,Fenz2011}. Numerous studies have explored the effect of applied forces  \cite{Guttenberg2000,Prechtel2002,Smith2007b,Smith2004,Smith2006,Shi2006,Cheng2009,Freund2009}, substrate curvatures \cite{Rosso1998,Shi2006,Das2008,Agrawal2009} and substrate elasticity \cite{Agrawal2011,Yi2011} on the equilibrium state of vesicles. Going beyond quasistatic models, dynamic rearrangement of adhesion molecules and adhesion domains have been studied in \cite{Smith2005a,Reister-Gottfried2008,Freund2004,Shenoy2005,Smith2008a,Fenz2011}. For a detailed review on adhesion, we refer the reader to the excellent reviews in  \cite{Smith2007a,Smith2009,Weikl2009,Sackmann2014a}. A few techniques for estimation of adhesion energy based on vesicle morphology have already been proposed in the literature. A contour analysis method relying on curvature and angle measurements near the contact boundary was developed by Sackmann and co-workers \cite{Albersdorfer1997, Guttenberg2000}. A comparison of the overall shape of the simulated and imaged vesicle was used in \cite{Smith2006}. Dynamic measurements of adhesion front were used by Bernard et al. \cite{Bernard2000}. Radii of the contact domain and the mid-plane of the adhered vesicle were used by Lai et al.  \cite{Lai2002}. An analytical functional relying on adhesion potential range, vesicle area and volume, contact area and spherical cap geometry was proposed by Lipowsky and co-workers \cite{Gruhn2007}. 

In this Letter, we build upon these fundamental studies to propose novel `universal relationships' to tease out adhesion energy from experimental measurements. Compared to the existing techniques, these remarkably simple relationships significantly reduce data processing and computational effort needed to extract adhesion energy post experiments. We restrict ourselves to a continuum framework. While binding between the adhesion molecules originates from the molecular forces, we model adhesion via an effective coarse-grained binding affinity or adhesion energy $\sigma$ defined per unit area. Such an energy would be a function of the areal density of these molecules and can enable a reasonably accurate investigation of force transmitted between cells without becoming overwhelmed by the molecular details. This approach has been successfully employed to investigate adhesion of membranes and vesicles (for example, \cite{Seifert1990,Lipowsky1991,Albersdorfer1997,Smith2007b,Deserno2007,Agrawal2009} to list a few). 


We model the equilibrium shape of a vesicle adhered to a planar rigid substrate. We model a bilayer as a 2D fluid surface that offers flexural resistance. The strain energy is assumed to be given by the Helfrich-Canham bending energy $W=kH^2 + \bar{k} K$, where $H$ is the mean curvature, $K$ is the Gaussian curvature, and $\{k,\bar{k}\}$ are the bending moduli \cite{Canham1970,Helfrich1973,JamesT.Jenkins1977,Steigmann1999,Deserno2015}. We further assume that the bilayer does not undergo areal dilation. As a result the total free energy of the vesicle is given by 
\begin{equation}
E=\int_{\omega}[W(H,K)+\lambda]da  -pV -\int_{\omega_c}\sigma da,
\end{equation}%
where $\lambda$ and $p$ are the Lagrange-multiplier fields associated with the areal constraint and the volumetric constraints. The last term accounts for the adhesion energy and the integration is performed over the contact domain $\omega_c$. It should be noted that $\sigma$ can depend on the surface coordinates and undergo spatial variation \cite{Agrawal2009}. Minimization of the free energy yields the well known shape equation in the free domain 
\begin{equation}
k[\Delta H+2H(H^{2}-K)]-2\lambda H=p.
\end{equation}
The variation of the adhesion energy, as shown in \cite{Agrawal2009}, yields $
\dot{E}_{\Gamma}=-\int_{\partial \omega_c }\sigma \mathbf{t\cdot u}ds$, where $\mathbf{t}$ is the exterior normal to the contact boundary and $\mathbf{u}$ is the tangential variation. Following the procedure outlined in \cite{Agrawal2009}, the boundary terms arising from the free and the adhered domains result in a single non-trivial jump condition at the interface 
\begin{equation}
[H]^{2}=\sigma /k,
\end{equation}%
where $[\cdot]= (\cdot)_c - (\cdot)_f$ and $\sigma$ is the value of the adhesion energy at the contact boundary. This jump condition establishes a link between the mean curvatures in the adhered and free domains. For a planar substrate, $H_c=0$, and therefore, $H_f=\sqrt{\sigma/k}$.

For the second part of the study, we simulate the effect of a vertical force $F$ applied at the (north) pole of the adhered vesicle. We compute the critical force required to detach the vesicle from the substrate. As shown in \cite{Agrawal2009}, force equilibrium of a subdomain $\tilde{\omega}$ surrounding a pole can be expressed as 
\begin{equation}
\int_{\tilde{\omega}}p\mathbf{n}da+\int_{\partial \tilde{\omega}}\mathbf{f}%
dt+F\mathbf{k=0,}
\end{equation}%
where $p$ is the pressure across the bilayer and $\textbf{f}$ is the force per unit length exerted on $\tilde{\omega}$ by the part $\omega \setminus \tilde{\omega}$. This yields the boundary condition $L=F/2\pi k$ at the pole, where
\begin{equation}
L=rH^{\prime} 
\end{equation}%
is the transverse shear force acting at a boundary and $(\cdot)'$ is the arclength derivative \cite{Agrawal2009}. 

We assume that the equilibrium shapes of the vesicles possess axisymmetry and set up the system in polar coordinates. We solve the shape equation [Eq. (2)], which reduces to
\begin{equation}
L^{\prime }=r[(2\lambda /k)H-2H(H-r^{-1}\sin \psi )^{2}],
\end{equation}%
in the free domain along with the geometric relations 
\begin{equation}
r^{\prime }(s)=\cos \psi, \quad z^{\prime }(s)=\sin \psi,
\end{equation}%
\begin{equation}
r\psi ^{\prime }=2rH-\sin \psi,
\end{equation}%
Eq. (7), and the boundary conditions to compute the equilibrium geometry of the vesicle. Above, $r$ is the radial distance from the axis of symmetry, $z$ is the height, $\psi$ is the angle the surface tangent makes with the horizontal, $s$ is the arclength, and $()'=d()/ds$.

The initial vesicle is assumed to possess a nearly spherical shape of radius $R$. This furnishes a lengthscale to non-dimensionalize the variables. In addition, we use the one-to-one correspondence between the arclength and the area ($da=2\pi r ds$) to switch to area as the independent variable. The constraint on the area is then simply imposed by integrating the equations over the desired area domain. The key dimensionless variables for this system are given by 
\begin{equation}
\alpha=a/2\pi R^2, \quad x=r/R, \quad y=z/R, \quad h=RH, 
\end{equation}%
\begin{equation}
\bar{\lambda}=R^2\lambda/k, \quad \bar{\sigma}=R^2\sigma/k, \quad \text{and} \quad f=FR/2 \pi k. 
\end{equation}%
We do not reproduce the normalized equations for the sake of brevity. However, these equations and the associated details can be found in \cite{Agrawal2009}. 

We simulate the equilibrium shape of a vesicle for a range of prescribed adhesion energies. As suggested by Lipowsky and Seifert \cite{Lipowsky1991}, we assume that the vesicles have remained adhered for a long time, letting the water permeate across the membrane allowing them to attain an equilibrium configuration for nearly zero transmembrane pressure conditions. We solve the shape equation in the free domain subject to the boundary conditions at the pole and at the contact interface. At the pole $x=0, \psi=0$. At the contact boundary, $y=0$, $x=r_c/R$, $\psi=-\pi$,  and $h=\sqrt{\bar{\sigma}}$ (here, we have dropped the subscript $f$). We choose a sequence of contact areas, which determines $r_c$, and identify the shape for which $L=0$. This is the equilibrium shape for a given $\sigma$ with no force applied acting on the vesicle. For the computed geometries for different $\sigma$, we record the vesicle height $y_p$ and the contact area $\alpha_c=(r_c/R)^2$. 

\begin{figure}[b]
\begin{center} 
\includegraphics[height = 77mm, keepaspectratio]{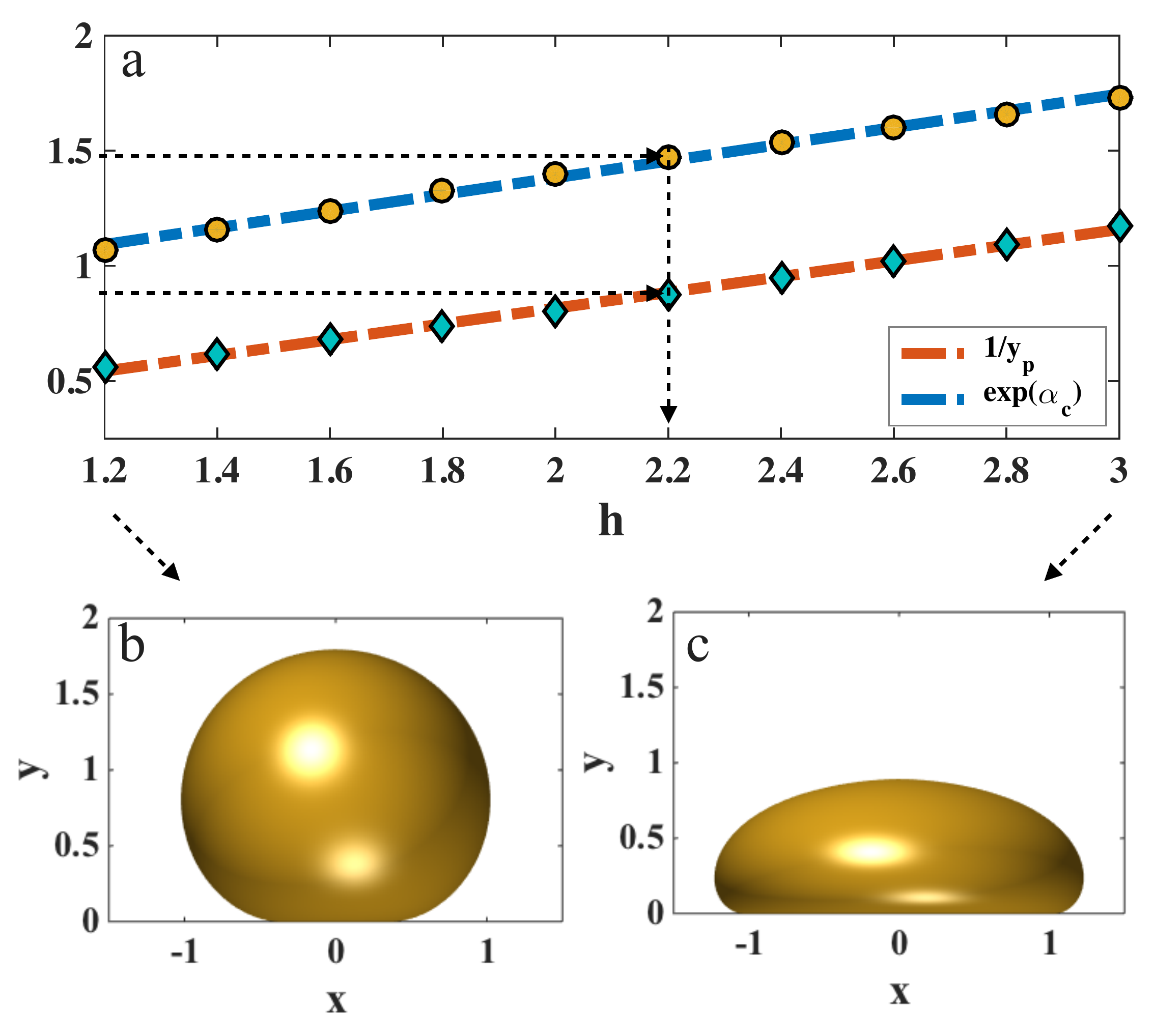} \hspace{-.5cm} \label{a}
\caption{a) Plots of $1/y_p$ and $exp(\alpha_c)$ as a function of the curvature of the free domain at the contact boundary ($h$). The two plots show a remarkable linear trend and furnish universal relationships to estimate adhesion energy from microscopy data. Measurement of any one of the two quantities can yield the corresponding value of $h^*$ (dashed lines) which can then provide an estimate of $\sigma^*$. b) Computed vesicle shape for $h=1.2$. c) Computed vesicle shape for $h=3.0$.}
\label{F2}
\end{center}
\end{figure}

Fig. \ref{F2}a shows the plots of the inverse of the normalized vesicle height ($1/y_p$) and the exponential of the contact area $exp(\alpha_c)$ as a function of the normalized interface curvature. Unexpectedly, what emerges are parallel straight lines that connect the computed data points. As the adhesion energy increases, the vesicle flattens out resulting in a reduction in the height and an increase in the contact area. Figs. 2b and 2c show the vesicle shapes for the two ends of the simulated domain. Since all the quantities are normalized, the linear relationships in Fig. \ref{F2}a are \textit{universal} and provide a means to estimate the adhesion energy. The profile of the adhered vesicle can be imaged with a microscopy technique that enables measurement of the height, contact radius, and the side profile of the adhered vesicle. With this data, one can compute the total surface area of the vesicle and the radius $R^*$ of the undeformed vesicle. The height of the vesicle and the contact area can then be normalized with $R^*$ to compute $1/y^*$ and $exp(\alpha_c^*)$. Either $1/y^*$ or $exp(\alpha_c^*)$ can then be used to read the normalized $h^*$ from Fig. \ref{F2} (black dashed arrows). Since the bending modulus of bilayers is known for a wide variety of lipids, the adhesion energy $\sigma^*$ can be computed as $k^*(h^*/R^*)^2$. Compared to the existing techniques, the proposed approach relies on fewer and simpler measurements (such as vesicle height and contact area) that might provide a more efficient way to estimate adhesion energy.

\begin{figure}[b]
\begin{center} 
\includegraphics[height = 76mm, keepaspectratio]{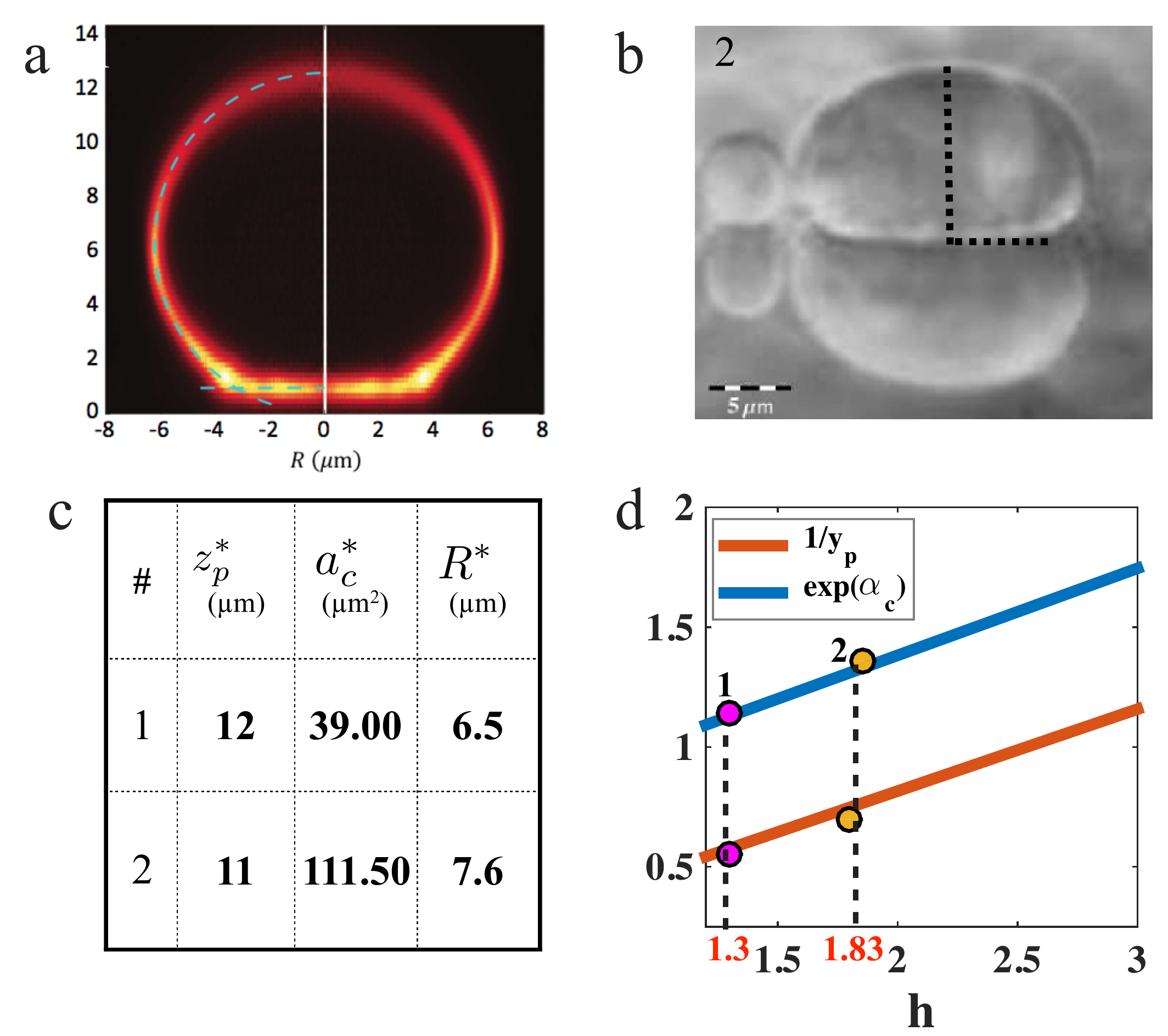} \hspace{-.5cm} \label{a}
\caption{Validation of modeling predictions with existing data on vesicle adhesion. a) Profile of an adhered vesicle reconstructed from confocal microscopy measurements \cite{Ursell2009}. Image reproduced with permission from T.U. b) Image of an adhered vesicle measured in \cite{Gruhn2007}. c) Vesicle height and contact area extracted from (a) and (b) d) As predicted by the simulations, normalized data in (c) lie on the predicted linear plots and yield $h^*_1=1.3$ and $h^*_2=1.83$. }
\label{F3}
\end{center}
\end{figure}


We now use experimental data from two previous studies to validate the proposed methodology. The first data (Fig. \ref{F3}a) is taken from \cite{Ursell2009}. Giant unilamellar vesicles and substrate were coated by a protein L1 implicated in neuronal cell growth. Confocal microscopy was used to read the circular cross sections of the adhered vesicle at different heights, which were then used to reconstruct the side profile of the vesicle shown in Fig. \ref{F3}a. The vesicle profile in Fig. \ref{F3}a yields $z_p^*=  12 \mu{m}$, $a_c*= 39 \mu{m}^2$, and $R^*= 6.5 \mu{m}$ (Table 1, Fig \ref{F3}c). These together yield $1/y_p^*= 0.54$ and $exp(\alpha_c^*)= 1.15$. The second set of data (Fig. \ref{F3}b) is taken from \cite{Gruhn2007}. Giant unilamellar vesicles were allowed to adhere to pure glass substrates and optical microscopy in phase contrast mode was employed to image the vesicle. The vesicle profile in Fig. 3b yields $z_p^*=  11 \mu{m}$, $a_c*= 111.5 \mu{m}^2$, and $R^*= 7.6 \mu{m}$ (Table 1, Fig 3c). These together yield $1/y_p^*= 0.69$ and $exp(\alpha_c^*)= 1.36$. Fig. \ref{F3}d shows the normalized experimental data on the predicted universal plots. As predicted by simulations, the two sets of data points corresponding to $1/y_p^*$ and $exp(\alpha_c^*)$ fall on the same vertical line yielding a unique $h^*$. For the first data set, $h^*=1.3$. Assuming a bending modulus of $80 k_BT$ (a value used in \cite{Ursell2009}), $h^*=1.3$ yields an adhesion energy $\sigma^*= 1.4\times10^{-8}J/m^{2}$. For the second data set, $h^*=1.83$, which yields $\sigma^*= 2.4\times10^{-9}J/m^{2}$ (for $k\approx 10 k_BT$ \cite{Gruhn2007}), a value in close agreement with an estimate of $10^{-9}-10^{-8}J/m^{2}$ in \cite{Gruhn2007}. 

\begin{figure}[b]
\begin{center} 
\includegraphics[height = 78mm, keepaspectratio]{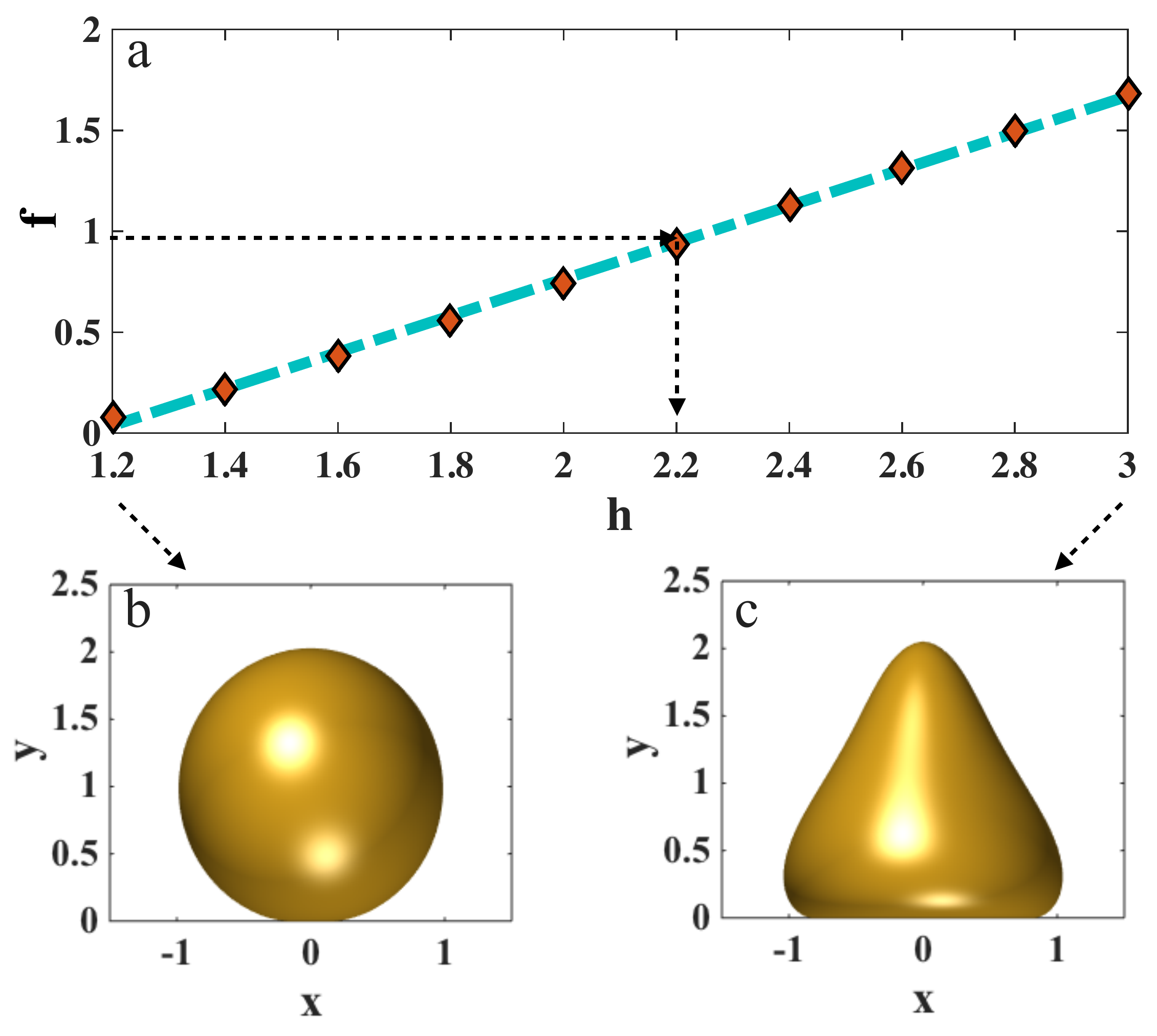} \hspace{-.5cm} \label{a}
\caption{a) Critical force $f$ required to pull the vesicle off the substrate as a function of $h$. The plot shows a simple linear trend and provides another universal relationship for computation of the adhesion energy. Experimental measurement of $f^*$ can directly provide an estimate $h^*$ (dashed lines) that will yield the value of $\sigma^*$. b) Computed vesicle shape at the critical force for $h=1.2$. c) Computed vesicle shape at the critical force for $h=3.0$.}
\label{F4}
\end{center}
\end{figure}

We now present another methodology for estimating adhesion energy. Instead of relying on the vesicle shape, this approach relies on measuring the vertical point force (applied at the north pole of the vesicle) required to overcome vesicle-substrate adhesion and pull the vesicle off the surface. Such a force, in principle, could either be applied by a magnetic bead or an AFM tip. Such techniques have been previously employed for physical manipulation of vesicles and membranes \cite{Guttenberg2000,Sen2005}. To simulate this scenario, we apply an upward-acting point load at the upper pole of the vesicle. As the force is increased, the vesicle begins to debond form the substrate. We ramp up the force until the vesicle detaches from the substrate. We record this critical force $F$ and normalize it to compute $f=FR/2\pi\kappa$. We repeat these simulations for a range of adhesion energies and plot them in Fig. \ref{F4}a. The vesicle shapes at the two ends of the simulated domain are shown in Figs. \ref{F4}b and \ref{F4}c. As before, the normalized critical debonding force $f$ in Fig. \ref{F4}a linearly increases with $h$ and again yields a universal relationship providing another approach to measure adhesion energy. Thus, if the critical force $F^*$ required to detach the vesicle is measured in an experiment, we can compute the normalized force $f^*=FR^*/2\pi\kappa$ and read the corresponding $h^*$ from Fig. \ref{F4}a (dashed lines). As before, with the knowledge of the bending modulus of the bilayer, $h^*$ can then be used to estimate adhesion energy $\sigma^*=k^*(h^*/R^*)^2$. 



As mentioned earlier, there exists a rich literature on membrane adhesion. In this study, we have resorted to a simplified but widely used mathematical framework. One specific phenomenon not accounted for in the current work is the impact of receptor mobility on vesicle adhesion. While we suppress receptor dynamics, our model does entertain spatial heterogeneity in the adhesion energy (see Eq 1). As shown in \cite{Agrawal2009}, equilibrium shape of the vesicle is then solely determined by the adhesion energy at the contact boundary and is insensitive to the variation inside the contact domain. Thus, for a system with mobile receptors, the first methodology will yield the adhesion energy at the boundary arising from the equilibrated distribution of receptors and will still maintain its applicability. In contrast, the second methodology based on the detachment force is limited to the case of immobile receptors. This occurs because the present study does not account for the force-induced rearrangement of receptors revealed in \cite{Smith2008a}. However, if the receptor density is high, force-induced redistribution can be expected to be less dominant and the proposed strategy should remain valid. Finally, the application of force on strongly adhered vesicles has been shown to lead to tubule formation \cite{Smith2004}. Since we do not encounter tubules in our simulations, our results are applicable in the ultra-weak to weak adhesion regimes. 

The results presented in this Letter may have broad implications in the areas of biophysics and biomaterials. The universal relationships presented in Figs. \ref{F2} and \ref{F4}  significantly reduce the need for post-processing of experimental data and do not require new simulations post-experiments to calibrate adhesion energy. In addition, these relationships can enable scientists to use a wider variety of optical and force microscopy techniques to measure adhesion energy. The same relationships can also be employed for surface characterization of elastic and biomimetic nanoparticles. It is important to note that an axisymmetric elastic nanoparticle made of a synthetic material is indistuingashable from an axisymmetric vesicle, as both behave as soft elastic shells. Any difference that may arise from the solid-like response of the synthetic material is suppressed as the off-diagonal terms in the metric tensor vanish in the axisymmetric setup\cite{Agrawal2009}. Since the adhesive properties of nanoparticles greatly influence their interactions with cells \cite{Samir2008,Ferrari2008Biomaterial,Zhang2010PRL,Bao2015ACSNANO}, universal relationships can provide a means for rationally functionalizing their surfaces. 

While a priori knowledge of bending stiffness enables estimation of adhesion energy, a priori knowledge of adhesion energy can enable estimation of bending stiffness of a vesicle or an elastic nanoparticle. Thus, if the adhesion strength is known for a given set of adhesion molecules or a material-substrate pair, the height or the contact area of an adhered nanoparticle can be used to compute the bending modulus $k^*=\sigma^*(R^*/h^*)^2$ from Fig. \ref{F2}. This method provides a novel technique for characterization of the elastic properties of nanoparticles which critically impacts their biological response \cite{Samir2008,Samir2015,Bao2015ACSNANO}. A subtle but important remark is in order here. Experimental techniques such as atomic force microscopy (AFM), commonly used to measure elastic property, provide an estimate of the Young's modulus of a nanoparticle. In contrast, the modeling studies which analyze cellular uptake of elastic nanoparticles require a knowledge of their bending modulus \cite{Gao2011,Bao2015ACSNANO}. Thus, the material property measured from the experiments and that used by the  biophysical models lack compatibility. The proposed technique based on nanoparticle adhesion can address this issue and provide a direct measurement of the bending modulus. Furthermore, it is important to note that Young's modulus of an elastic nanaoparticle inferred from AFM data is not a pure reflection of the material property. It also imbibes in it the stiffness imparted by the shell-like geometry. As a consequence, AFM data for two spherical nanoparticles made of the same material and thickness but with different radii would yield two different Young's moduli. However, from the point of view of curvature elasticity, the two nanoparticles have identical material property (same bending modulus). Thus, the use of Young's modulus makes it challenging to disentangle the roles of elasticity and geometry in nanoparticle-cell interactions, which have distinct impacts on the fate of nanoparticles \cite{Samir2008,Gao2011,Samir2015,Mitragotri2006PNAS,Suresh2009AdvanceMaterials,Zhang2013Nanoletter,Bao2015ACSNANO}.

In summary, in this Letter, we simulated vesicle-substrate interactions and presented three universal relationships. The inverse of the vesicle height, exponential of the contact area, and the critical detachment force all vary linearly with the square root of the adhesion energy. Despite the fact that vesicle adhesion has been extensively studied over the last few decades, the existence of such universal relationships has remained elusive to date. These relationships can potentially open new avenues for biophysical characterization of adhesion molecules and elastic nanoparticles.



\bibliographystyle{apsrev}
\bibliography{MyCollection}{}

\end{document}